\documentclass[%
 aip,
 amsmath,amssymb,
 reprint,%
floatfix]{revtex4-1}

\usepackage{graphicx}
\usepackage{dcolumn}
\usepackage{bm}

\usepackage[utf8]{inputenc}
\usepackage[T1]{fontenc}
\usepackage{mathptmx}
\usepackage{etoolbox}
\usepackage{siunitx}
\usepackage{upgreek}
\usepackage{placeins}

\makeatletter
\def\@email#1#2{%
 \endgroup
 \patchcmd{\titleblock@produce}
  {\frontmatter@RRAPformat}
  {\frontmatter@RRAPformat{\produce@RRAP{*#1\href{mailto:#2}{#2}}}\frontmatter@RRAPformat}
  {}{}
}%
\makeatother

\begin{document}

\preprint{AIP/123-QED}

\title{Frequency fluctuations of ferromagnetic resonances at milliKelvin temperatures}
\author{Tim Wolz}
 	\affiliation{Institute of Physics, Karlsruhe Institute of Technology, 76131 Karlsruhe, Germany}
\author{Luke McLellan}
	\affiliation{James Watt School of Engineering, Electronics \& Nanoscale Engineering Division, University of Glasgow, Glasgow G12 8QQ, United Kingdom}
\author{Andre Schneider}
	\affiliation{Institute of Physics, Karlsruhe Institute of Technology, 76131 Karlsruhe, Germany}
\author{Alexander Stehli}
\affiliation{Institute of Physics, Karlsruhe Institute of Technology, 76131 Karlsruhe, Germany}
\author{Jan David Brehm}
\affiliation{Institute of Physics, Karlsruhe Institute of Technology, 76131 Karlsruhe, Germany}
\author{Hannes Rotzinger}
\affiliation{Institute of Physics, Karlsruhe Institute of Technology, 76131 Karlsruhe, Germany}
\affiliation{Institute for Quantum Materials and Technologies, Karlsruhe Institute of Technology, 76131 Karlsruhe, Germany}
\author{Alexey V. Ustinov}
\affiliation{Institute of Physics, Karlsruhe Institute of Technology, 76131 Karlsruhe, Germany}
\affiliation{National University of Science and Technology MISIS, 119049 Moscow, Russia}
\affiliation{Russian Quantum Center, 143025 Skolkovo, Moscow, Russia}
\author{Martin Weides}	
\email{martin.weides@glasgow.ac.uk}
\affiliation{James Watt School of Engineering, Electronics \& Nanoscale Engineering Division, University of Glasgow, Glasgow G12 8QQ, United Kingdom}

\date{\today}

\begin{abstract}
Unwanted fluctuations over time, in short, noise, are detrimental to device performance, especially for quantum coherent circuits. Recent efforts have demonstrated routes to utilizing magnon systems for quantum technologies, which are based on interfacing single magnons to superconducting qubits. However, the coupling of several components often introduces additional noise to the system, degrading its coherence. Researching the temporal behavior can help to identify the underlying noise sources, which is a vital step in increasing coherence times and the hybrid device performance. Yet, the frequency noise of the ferromagnetic resonance (FMR) has so far been unexplored. Here, we investigate such FMR frequency fluctuations of a YIG sphere down to mK-temperatures, and find them independent of temperature and drive power. This suggests that the measured frequency noise in YIG is dominated by so far undetermined noise sources, which properties are not consistent with the conventional model of two-level systems, despite their effect on the sample linewidth. Moreover, the functional form of the FMR frequency noise power spectral density (PSD) cannot be described by a simple power law. By employing time-series analysis, we find a closed function for the PSD that fits our observations. Our results underline the necessity of coherence improvements to magnon systems for useful applications in quantum magnonics.
\end{abstract}

\maketitle

Fluctuations of the resonance frequency and other forms of noise can drastically hamper the performance of sensors, amplifiers, and information processing circuits. This is accurate at room temperature but particularly crucial for quantum devices, where environmental noise leads to decoherence. With the recent coupling of single magnons to superconducting qubits\cite{tabuchi_coherent_2015, lachance-quirion_resolving_2017, lachance-quirion_entanglement-based_2020} and resonators\cite{morris_strong_2017, li_strong_2019, baity_strong_2021}, research on hybrid quantum magnonics\cite{lachance-quirion_hybrid_2019, li_hybrid_2020, elyasi_resources_2020} has emerged. There, the goal is a combination of quantum computing's exponential speed-up with magnonics'\cite{kruglyak_magnonics_2010, chumak_magnon_2015} low-loss devices. First demonstrations of magnonic devices are, for example, a magnon based transistor\cite{chumak_magnon_2014} or a majority gate\cite{klingler_design_2014}, combining OR and AND logic. Moreover, with a radio frequency-to-light conversion based on magnons\cite{hisatomi_bidirectional_2016, osada_cavity_2016, zhang_optomagnonic_2016}, a possible direction towards a quantum internet exists, but also requires a coupling of several quantum systems. Such a coupling often gives rise to additional loss channels and increased noise, which along with the short coherence times of magnons presents a major obstacle in quantum magnonics\cite{clerk_hybrid_2020}. Yet, the influence and origin of magnonic noise is still largely an open question. Predominantly, phase noise has been considered in magnetic tunnel junction oscillators \cite{houssameddine_temporal_2009, quinsat_amplitude_2010}, the amplitude noise in a magnonic waveguide \cite{rumyantsev_discrete_2019} at room temperature, and theoretically the magnetization noise of spins\cite{kubo_brownian_1970, miyazaki_brownian_1998, foros_noise_2009}, for instance. Frequency fluctuations of the most basic magnon mode, the ferromagnetic resonance (FMR), however, have eluded attention.

Here, we experimentally observe such FMR frequency fluctuations with a focus on an yttrium-iron-garnet (YIG) sphere at mK temperatures and show that time-series analysis can yield additional information, especially when the noise frequency dependence of the fluctuations cannot be described by a simple power law. After an introduction to the measurement setup and the spectroscopic characterization of the YIG sample, we briefly recapitulate the concept of the power spectral density (PSD). Then, the results of the frequency noise measurements are presented and analyzed. After which, we compare the results to room temperature data and a different material, lithium ferrite (LiFe).

\begin{figure*}
    \centering
    \includegraphics{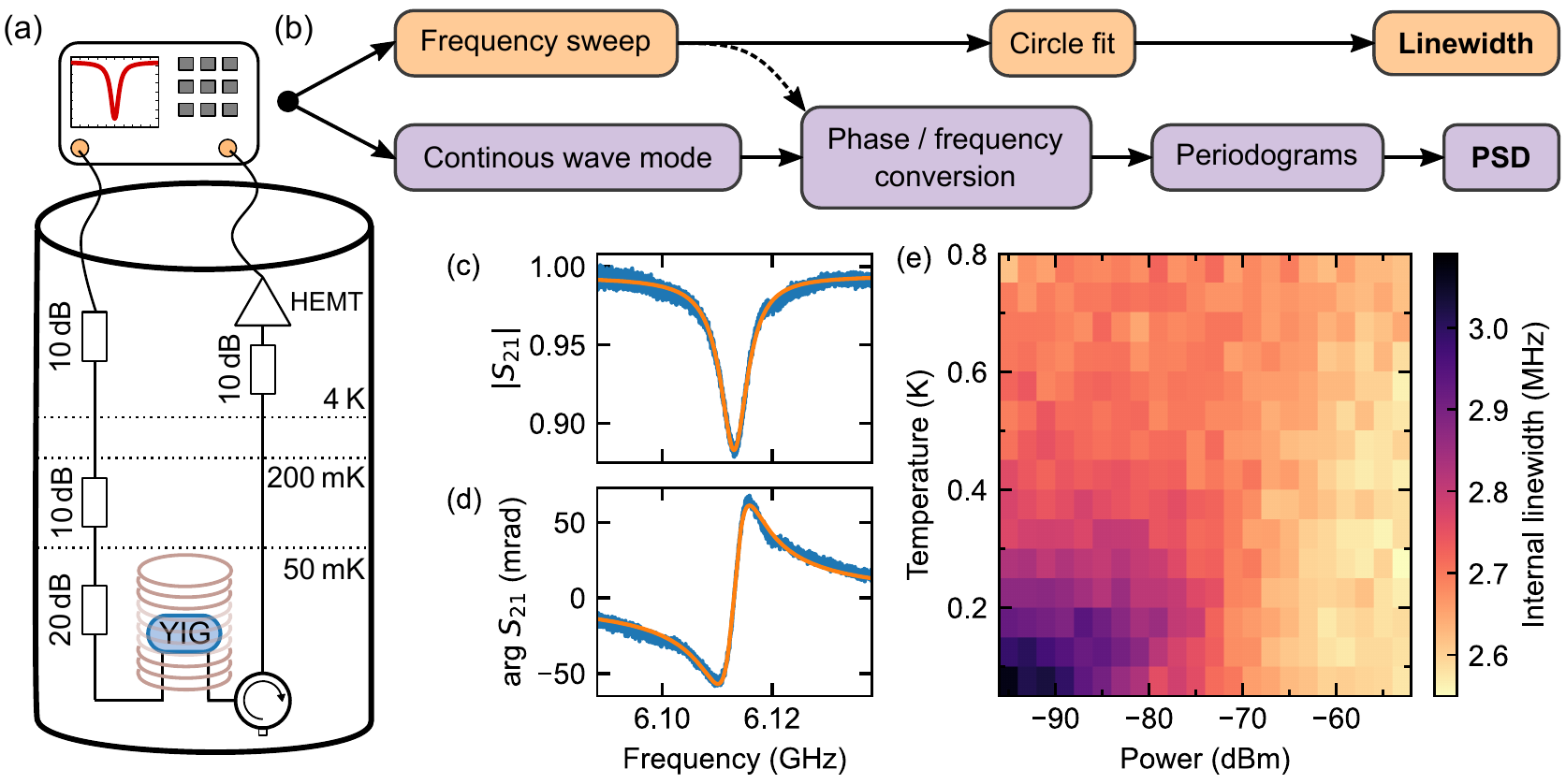}
    \caption{Experimental setup, measurement schemes, and sample characterization at mK temperatures. (a) The magnetic medium, a YIG sphere, is mounted over a micro strip line and placed inside a solenoid coil in a dilution refrigerator. The steady-state response is measured via a vector network analyzer. (b) The measurement schemes show how linewidth data is extracted from frequency sweeps, whereas the continuous wave mode allows for an estimation of the frequency noise power spectral density (PSD).  (c,d) Amplitude $|S_{21}|$ and phase $\arg S_{21}$ response of the ferromagnetic resonance, shown for input power at the sample of $P=\SI{-90}{dBm}$ and temperature $T=\SI{50}{\milli \kelvin}$ (background corrected). Solid orange lines denote a circle fit\cite{probst_efficient_2015}, which is used to determine the FMR linewidth. Linear region of the phase response yields the conversion from phase fluctuation to frequency fluctuations. (e) Internal linewidth extracted from circle fits shows a temperature and power dependence that was previously attributed to loss into a bath of two-level systems\cite{tabuchi_hybridizing_2014, kosen_microwave_2019, pfirrmann_magnons_2019}.}
    \label{fig:setup_charac}
\end{figure*}

Our experimental setup (Fig.~\ref{fig:setup_charac}\,(a)) consists of a vector network analyzer (VNA) connected to the different magnetic media via a strip-line in a notch-type configuration. For the mK temperature measurements, the sample, a YIG sphere with diameter $d=\SI{0.2}{\milli \meter}$, is mounted in a solenoid coil inside a dilution refrigerator. A VNA offers a straight forward procedure for frequency noise measurements. Sweeping the probe frequency allows for a characterization of the sample via its $S_{ij}(\omega)$-matrix element, from which we extract the FMR linewidth. Then, to measure frequency fluctuations, we employ the continuous wave mode of the VNA with probe frequency $\omega_\mathrm{p}$. Here, we record a time trace of the sample's frequency response at one single point close to resonance ($\omega_\mathrm{p} \approx \omega_\mathrm{r}$). Fluctuations in the phase $\arg S_{21}$ can then be converted to resonance frequency fluctuations via the slope in the linear region of $\arg S_{21}(\omega)$, see Fig.~\ref{fig:setup_charac}\,(b) for a schematic overview and Supplementary Information~\ref{sec:exp_detail} for more details. All measurements are performed and evaluated with the open-source measurement suite qkit\cite{qkitgroup_qkit_nodate}.

We start with the spectroscopic characterization of our sample at mK temperatures. The FMR is tuned to $\omega_\mathrm{r}/2\pi=\SI{6.11}{\giga \hertz}$, corresponding to an external field $\upmu_0 H \approx \SI{0.21}{\tesla}$, where the sample is fully magnetized. Figures~\ref{fig:setup_charac}\,(c,d) show the amplitude and phase of the background corrected complex $S_{21}$ frequency response. A circle fit\cite{probst_efficient_2015} returns the internal linewidth (HWHM) $\kappa_{\mathrm{i}} = \omega_\mathrm{r}/(2 Q_i)$, with $Q_\mathrm{i}$ as internal Q-factor. Varying power and temperature, we find a linewidth dependence that decreases with increasing power $P$ and temperature $T$ in accordance to previous reports, which attributed this effect to energy loss into a bath of two-level systems (TLS) \cite{tabuchi_hybridizing_2014, kosen_microwave_2019, pfirrmann_magnons_2019} (see Fig.~\ref{fig:setup_charac}\,(e)). Increasing temperature and power eliminates the loss channels into the TLS bath by equalizing the occupation numbers of excited and unexcited states of the TLS bath. In the standard tunneling model, this linewidth dependence is given by \cite{muller_towards_2019}

\begin{equation}
    \kappa_{\mathrm{TLS}} \propto \frac{\tanh\left({\hbar \omega_\mathrm{r}/k_\mathrm{B}T}\right)}{\sqrt{1+P/P_\mathrm{c}}}.
\end{equation}

$P_\mathrm{c}$ denotes the critical drive power, at which the Rabi drive rate exceeds the coherence of the TLS. For our sample $P_\mathrm{c}$ can be found in the range of $\SI{-80}{dBm}<P_\mathrm{c}<\SI{-70}{dBm}$. The  cable loss is included and estimated to be \SI{20}{\decibel}. These values correspond to photon numbers of $10^6$ to $10^7$ and are similar to previous results for magnon excitations in YIG\cite{kosen_microwave_2019, pfirrmann_magnons_2019}.

We now focus on the noise measurements. A recorded time trace of a fluctuating parameter, in our case frequency fluctuation $\Delta f$, can be difficult to interpret and, hence, the power spectral density $S(f)$ of the underlying random process is estimated. Roughly speaking, the PSD represents the fluctuation strength for a given frequency interval. The Wiener Khinchin theorem \cite{wiener_generalized_1930} relates the autocorrelation function (ACF) of the measured time trace to the PSD via Fourier transform. Employing the convolution theorem, one can calculate a so-called periodogram, an estimate of the PSD:
\begin{align}
S_{\Delta f}(f) &= \lim \limits_{T\to \infty} \frac{1}{2T}|\mathcal{F}(\Delta f_\mathrm{r}(t))|^2 \nonumber \\
&\approx \frac{1}{N\, f_\mathrm{s}} \left|\sum_{n=1}^N \Delta f_{\mathrm{r},n} \, \mathrm{e}^{-\mathrm{i} 2\pi f n \delta t}\right|^2.
\label{eq:periodogram}
\end{align}
In the second line, we used the discrete Fourier transformation with $N$ data points, sampling time $\delta t$ and normalization by the sampling rate $f_\mathrm{s}=1/\delta t$. For clarity, a subscript $r$ for the frequency fluctuations $\Delta f$ is added in these equation. To reduce the PSD's variance, we utilize Welch's method \cite{welch_use_1967}, where the data is divided into multiple segments and the resulting periodograms are averaged.

\begin{figure*}
    \centering
    \includegraphics{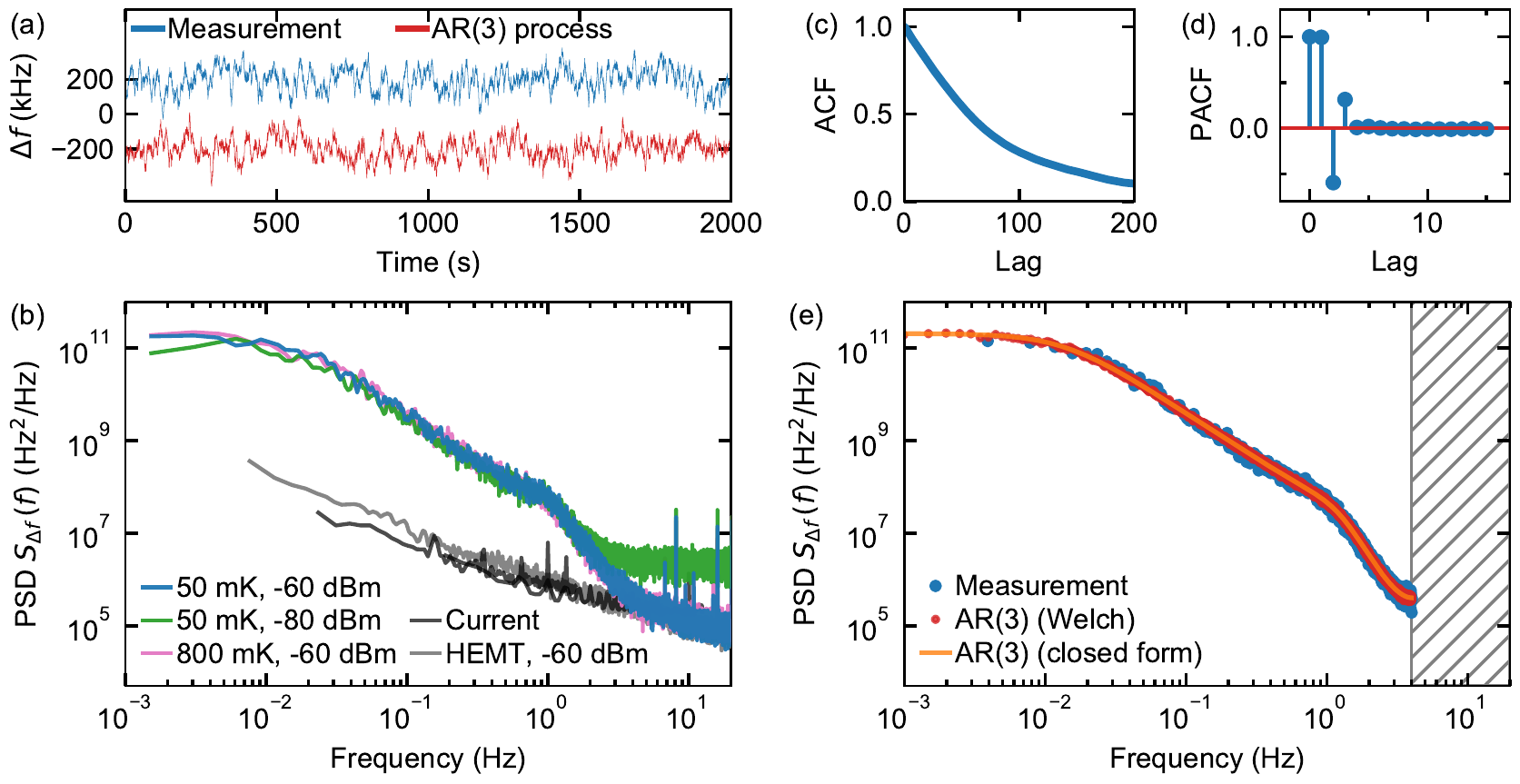}
    \caption{Frequency fluctuations of a ferromagnetic resonance (FMR) at mK temperatures. (a) Measured time trace of the frequency fluctuations ($P=\SI{-60}{dBm},\:T=\SI{50}{\milli \kelvin}$) compared to generated data, a realization of a third order autoregressive process AR(3). For better visibility the data are offset by $\pm \SI{200}{\kilo \hertz}$. Measured data is post averaged to \SI{8}{\hertz}, so that periodic signals are removed. (b) Power spectral density (PSD) of FMR frequency fluctuations for different input powers and temperature. Below \SI{3}{\hertz}, the sample noise PSD exceeds the amplitude of parasitic noise sources, such as the HEMT amplifier and the current source. No dependence on these external parameters can be observed in the low frequency range of the PSD. The functional form of the PSDs is close to a Lorentzian but shows a steeper decay at around \SI{1}{\hertz}. (c,d) Autocorrelation (ACF) and partial autocorrelation (PACF) of the time trace data displayed in (a). PACF only shows values significantly different from zero up to lag $n=3$, indicating an AR(3) process. Note the different scales of the x-axes. (e) A comparison of measured FMR noise data with an AR(3) process shows excellent agreement. For the generated data, the PSD is calculated via Welch's method and for the closed form PSD data, Eq.~(\ref{eq:PSD_ARp}) was employed with estimated coefficients.}
    \label{fig:PSDS}
\end{figure*}

Different physical noise mechanisms can manifest in distinct noise PSDs and are affected differently by external parameters. The presence of TLS does not only lead to an increased energy loss but TLS near resonance are also responsible for frequency fluctuations. However, these fluctuations can be covered by other, more dominating, noise sources. TLS frequency fluctuations can be understood in the Jaynes-Cummings model, where the resonance frequency of the FMR receives a shift depending on the state of the TLS. Such fluctuations have been observed in superconducting resonators\cite{gao_noise_2007, kumar_temperature_2008, neill_fluctuations_2013, burnett_evidence_2014, brehm_transmission-line_2017}, revealing three main characteristics: First, the frequency dependence of the PSD shows the infamous $1/f$ decay, which is explained\cite{muller_towards_2019, dutta_low-frequency_1981} by TLS uniformly distributed in frequency space with coherence rates distributed according to $P(\kappa) \propto 1/\kappa$. Second, due to the saturation of a TLS bath with power, the amplitude $A$ of the noise PSD should scale according to $A\propto (P/P_\mathrm{c})^{-0.5}$, as shown in  Ref.~\onlinecite{neill_fluctuations_2013}. Third, depending on whether the TLS themselves interact with each other, the amplitude should either reduce for decreasing temperature (non-interacting) or increase (interacting)\cite{burnett_evidence_2014}. We can now apply this knowledge to the results of our noise measurements, which are presented in Fig.~\ref{fig:PSDS}. Panel~(a) shows a recorded time trace of the FMR frequency fluctuations, which is used to evaluate the PSD, as displayed in Fig.~\ref{fig:PSDS}\,(b). We note that the observed frequency noise PSD is indeed higher than the amplifier noise and the noise produced by the current source (see Supplementary Information~\ref{sub:para_noise}). We also observe a functional form of the PSD that does not fit a simple power law. Up to \SI{1}{\hertz}, it can be described by a Lorentzian function but then a steep decrease follows. To test the influence of external parameters, we varied the temperature from \SIrange[]{50}{800}{\milli \kelvin} and swept the input power around the critical power from \SI{-60}{dBm} down to \SI{-100}{dBm}. Three curves are shown as examples, see Supplementary Information~\ref{sub:add_PSD} for more data. The frequency noise PSDs all show an independence of temperature and power. The increased white noise part for the low power PSD arises from amplifier noise. Taking all these points together, we conclude that TLS as described by the standard tunneling model are not the most dominant noise source for frequency fluctuations in our magnetic system. The lack of a power dependence is the strongest argument. A comparison to superconducting resonators\cite{burnett_evidence_2014} supports this statement. There, the TLS noise PSD at \SI{0.1}{\hertz} is three magnitudes lower than the observed FMR fluctuation PSD. Hence, despite showing a power or temperature dependent resonance linewidth, so far undetermined noise sources most likely mask the influence of TLS noise in the magnon system. 

As the measured PSD does not follow a simple power law, we search for a closed function that describes our data. For this purpose, we return back to the time trace and analyze it with a method closely related to maximum entropy spectral analysis \cite{bos_autoregressive_2002}, and based on time series analysis. There, a basic model describing random data is the autoregressive (AR) process, defined as
\begin{equation}
    y_t = \epsilon_t + \sum_{i=1}^{p} a_i y_{t-i}.
    \label{eq:AR}
\end{equation}
A random data point at time $t$ is calculated via a weighted sum of the last $p$ data points plus a white noise term with a Gaussian probability density function $\mathcal{N}(\sigma, \mu = 0)$, where $\sigma$ and $\mu$ are the standard deviation and mean value, respectively. The $a_i$ are free parameters and have to be estimated as well as the order $p$ of the process. AR processes are applicable if the influence of a single perturbation propagates via sums of exponential decays or damped oscillations. A famous examples is the AR(1) process with $a_i=1$, describing a random walk or Brownian motion. A reduction of $a_1$ results in the damping of these fluctuations over time. See Supplementary Information~\ref{sub:examples_AR} for more examples and higher order processes. To first test the applicability of an AR process to our data, we look at the ACF, which indeed shows an exponential-like decay (see Fig.~\ref{fig:PSDS}\,(c)) for the measured FMR frequency noise, and hence points towards an AR process. Next, we estimate the order of the process, as well as the values of our coefficients. Here, we make use of the partial autocorrelation function (PACF), which only returns the direct correlation between data points, i.e., the indirect influence of data points lying between is switched off. Since per definition of the AR process, a direct influence only exists up to order $p$, we count the time lags that show a value significantly different from zero, and find $p=3$ (Fig.~\ref{fig:PSDS}\,(c)), confirming the validity of the AR model for our data. The $a_i$ coefficients can then be calculated by employing the Yule-Walker equations\cite{yule_method_1927, walker_periodicity_1931}, which relate the ACF to the $a_i$ (Supplementary Information~\ref{sub:YWE}). The estimated coefficients are $a_1=1.764, \: a_2=-1.079, \: a_3=0.309, \: \sigma=\SI{5.284}{\kilo \hertz}$. Note that the order and subsequently the coefficients depend on the chosen sampling rate. To remove periodic signals and the $1/f$ amplifier part, a digital post averaging to a sampling frequency of \SI{8}{\hertz} was performed, see Supplementary Information~\ref{sub:PACF_sampl} for different sampling rates. With the estimated values, we can generate a model time trace for comparison (Fig.~\ref{fig:PSDS}\,(a)) and calculate its PSD. Importantly, a closed form\cite{box_time_2015} for the PSD of an AR($p$) process exists that depends on the $a_i$ parameters and the variance $\sigma^2$ of the white noise part:
\begin{equation}
    S(f) = \frac{2 \sigma^2 \delta t}{\left|1-\sum_{k=1}^{p}a_k\mathrm{e}^{-i 2\pi k\, \delta t \, f}\right|}.
\label{eq:PSD_ARp}
\end{equation}
Figure~\ref{fig:PSDS}\,(e) shows an excellent agreement of the measured PSD with both, the numerical simulation and the closed form. Furthermore, from the estimated parameters and the ACF, we can conclude that FMR frequency fluctuations are exponentially damped out over time without showing an oscillating behavior. The higher order of the AR process indicates several noise mechanisms that occur on different timescales\cite{miyazaki_brownian_1998}, and therefore require a weighted sum of the last $p$ data points.

\begin{figure}
    \centering
    \includegraphics{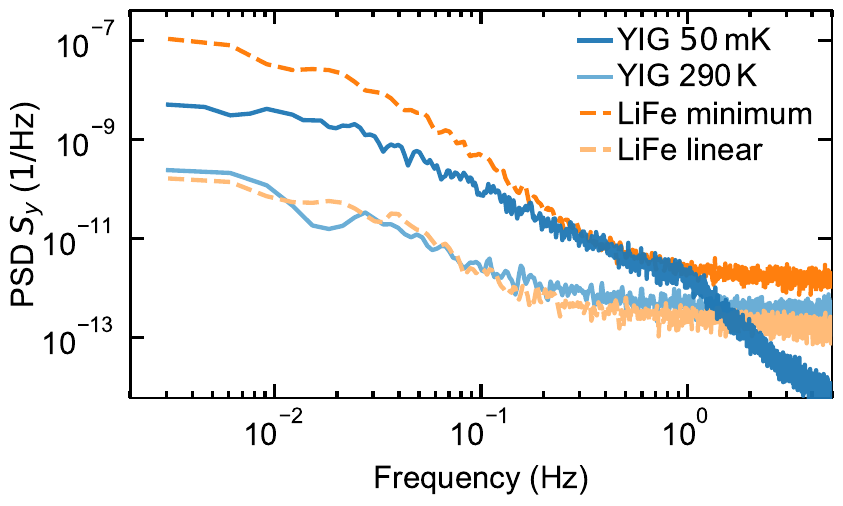}
    \caption{Comparison of frequency fluctuations of YIG and LiFe at room temperature to the low temperature data. The amplitudes of the power spectral densities (PSD) are normalized by the resonance frequency $S_y = S_{\Delta f}/f_\mathrm{r}^2$. LiFe exhibits a minimum in its field dispersion, resulting in a magnetic field insensitivity. The PSD at this point in the dispersion are compared to a point in the linear regime. Different white noise baselines are due to phase frequency conversion and normalization.}
    \label{fig:RT_PSD}
\end{figure}

We also compare the previous results to room temperature measurements of YIG and LiFe. Two surprising results can be observed from the data in Fig.~\ref{fig:RT_PSD}: First, focusing on YIG, we see that the frequency noise at low temperature is about two magnitudes higher than at room temperature in the low frequency region and both curves exhibit different functional forms. The increased white noise level, compared to the low temperature measurement, can be attributed to the lower dynamic range of a second VNA, employed in the room temperature setup. We note that these room temperature fluctuations could be caused by current noise, which we could not directly measure in this setup. Nevertheless, the higher amplitude at low temperature either suggests an extreme increase of noise with temperature in a region above \SI{800}{\milli \kelvin} to room temperature or distinctly different noise mechanisms. Both possibilities emphasize that additional care has to be taken in the development of coherent quantum magnonic devices in the future. Second, we examine frequency noise of LiFe. For this material, a mode softening was observed yielding a minimum in its dispersion \cite{goryachev_cavity_2018}. At such a minimum, the resonance frequency is first-order insensitive to field fluctuations. The FMR of our sample exhibits a gradient of the dispersion that is almost zero (see Supplementary Information~\ref{sub:RT_char}). We perform measurements in this region and in the linear dispersion regime. Despite the reduced field sensitivity, measurements at this insensitivity point show stronger fluctuations than in the linear dispersion regime. The fluctuations are also stronger than the low temperature noise of YIG. The strong frequency noise of LiFe around the insensitivity point therefore presents a considerable challenge for possible magnon based frequency precision applications\cite{flower_experimental_2019}. 

In conclusion, we studied FMR frequency fluctuations at mK temperatures. The recorded PSDs do not show a simple power law and are also independent of temperature and input power, which indicates undetermined noise mechanisms stronger than the influence of TLS described by the the standard tunneling model. We also presented a method to analyze noise data in the time domain, especially useful if a simple power law is not sufficient to describe the noise PSD. With this method and after post averaging of the data down to \SI{8}{\hertz}, we find an excellent agreement of the measured data with an AR(3) process, suggesting that several noise processes on different time scales are at play. A comparison to room temperature measurements and LiFe has shown increased noise at low temperatures and, surprisingly, also a high noise PSD close to the field-insensitivity point of LiFe, underpinning the importance of improving magnon coherence for useful applications. With this work, we hope to spark a broader interest into magnon decoherence research.

\begin{acknowledgments}
We wish to acknowledge fruitful discussions with Jürgen Lisenfeld, Khalil Zakeri, Dmytro Bohzko, and Mehrdad Elyasi. We acknowledge financial support from the former Helmholtz International Research School for Teratronics (Tim Wolz), the Landesgraduiertenf\"orderung (LGF) Baden-W\"urttemberg (Alexander Stehli), the Carl-Zeiss-Foundation (Andre Schneider) and Studienstiftung des Deutschen Volkes (Jan David Brehm). This work was supported by the European Research Council (ERC) under the Grant Agreement 648011 (MW) and  by the Ministry of Science and Higher Education of the Russian Federation in the framework of the State Program (Project No. 0718-2020-0025) (AVU).
\end{acknowledgments}

\section*{Data Availability Statement}
The data that support the findings of this study are available from the corresponding author upon reasonable request.

\bibliography{bib}

\clearpage
\newpage

\appendix
\section{Supplementary information}
\subsection{Experimental details}
\label{sec:exp_detail}
\subsubsection{Sample geometry}
\label{sub:sample_geom}
In the low temperature experiment, the YIG sphere is placed over a \SI{50}{\ohm} matched micro strip line such that the 110 axis is aligned parallel to the external field. The micro strip is made from a Rogers TMM10i copper cladded (\SI{35}{\micro \meter}) substrate with a thickness of \SI{0.64}{\milli \meter}.

For the room temperature experiments, the external field is generated by two Helmholtz coils with an iron yoke. The sample is also placed over a micro strip made from a Rogers TMM10i substrate with the 110 axis along the external field.

\subsubsection{Phase frequency conversion}
Phase fluctuations can be converted to frequency fluctuations via the following formula:
\begin{equation}
\Delta \varphi = \left(-2 Q_\mathrm{L} +2 Q_\mathrm{i} \right) \frac{\Delta \omega}{\omega_0}.
\label{eq:phase_to_freq}
\end{equation}
This equation represents the linearization of the phase roll-off around the resonance frequency. In our experiments, however, we fit the phase response around the resonance frequency by a linear function, and use the extracted parameters for the phase-frequency conversion. We also note that by having chosen sampling rates $f_\mathrm{s}<\SI{200}{\hertz}$ all noise measurements are performed below the Leeson frequency $\omega_\mathrm{L}=\omega_\mathrm{r}/(2Q_\mathrm{L})$, with $Q_\mathrm{L}$ as the loaded quality factor. This way only frequency fluctuations are observed and not instantaneous phase fluctuations \cite{rubiola_leeson_2005}. 

\subsubsection{Parasitic noise sources}
\label{sub:para_noise}
We identify two parasitic external noise sources in the setup: the HEMT amplifier, producing phase noise and the current source, with current fluctuations translating into frequency fluctuations of the FMR. Microwave amplifiers exhibit a low frequency $1/f$ part, which is independent of power, and a white noise part, scaling inversely scaling with input power\cite{boudot_phase_2012}. To reduce the amplifier's white noise, we prepared the sample in the under-coupled regime, so that most of the input power is transmitted. Yet according to Eq.~(\ref{eq:phase_to_freq}) the slope of the phase response flattens out and then again frequency noise of the sample will be low compared to the phase noise of the amplifier for a strongly under-coupled regime. As shown in Fig.~\ref{fig:setup_charac}\,(c,d) the amplitude and phase signal is still strong enough and hence a good compromise was found. Additionally, the sample is shielded from HEMT noise by a circulator and an additional \SI{10}{\decibel} attenuator before the HEMT, which is used to prevent compression of the HEMT due to the high input powers employed in the experiment. Band pass filters (\SIrange[range-units=single,range-phrase=--]{3}{7}{\giga \hertz}), installed before and after the sample, reduce unwanted external low frequency noise in the microwave lines.

To determine the current fluctuations, we inserted a \SI{1}{\ohm} resistor between current source and solenoid coil at room temperature. We then employed an FFT spectrum analyzer and measured the voltage drop at the resistor over an RC high pass with a cutoff frequency of $f_\mathrm{c}=\SI{3e-2}{\hertz}$ to filter out the dc part, thereby circumventing the dynamic range limitation of the spectrum analyzer.

\begin{figure}
    \centering
    \includegraphics{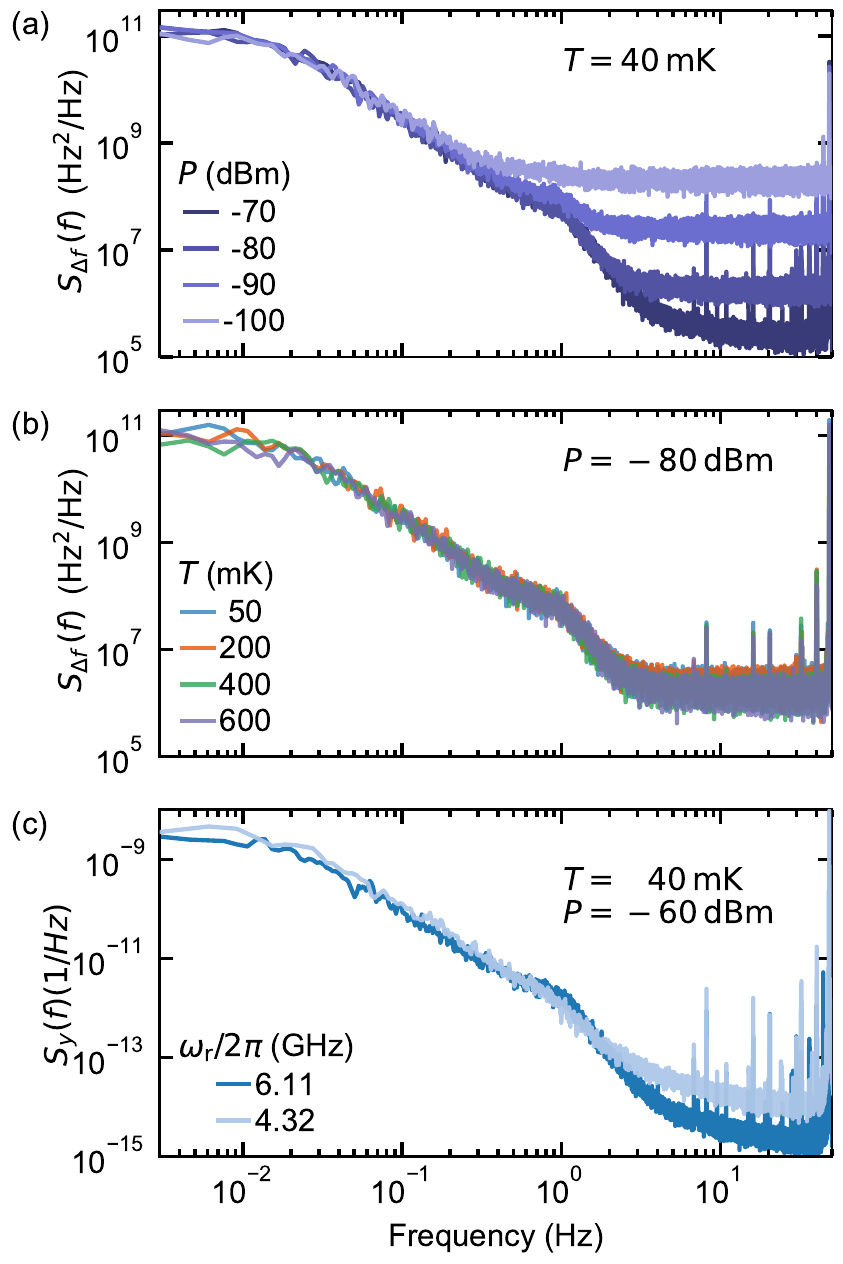}
    \caption{Measured power spectral densities for different power, temperature and resonance frequencies. (a) Power sweep, at constant temperature $T=\SI{40}{\milli \kelvin}$. Power is referenced to the sample input. No power dependence is visible. (b) Temperature sweep, at  constant power $P=\SI{-80}{dBm}$. All recorded PSDs are temperature independent. (c) Comparison of frequency fluctuations at two different resonance frequencies $\omega_{\mathrm{r},1} = \SI{6.11}{\giga \hertz}$ (used in (a) and (b)) and $\omega_{\mathrm{r},1} = \SI{4.32}{\giga \hertz}$. The PSDs are normalized to their resonance frequencies. Differences are minimal and only visible around the kink at \SI{1}{\hertz}. }
    \label{fig:psd_compar}
\end{figure}

\begin{figure*}
    \centering
    \includegraphics{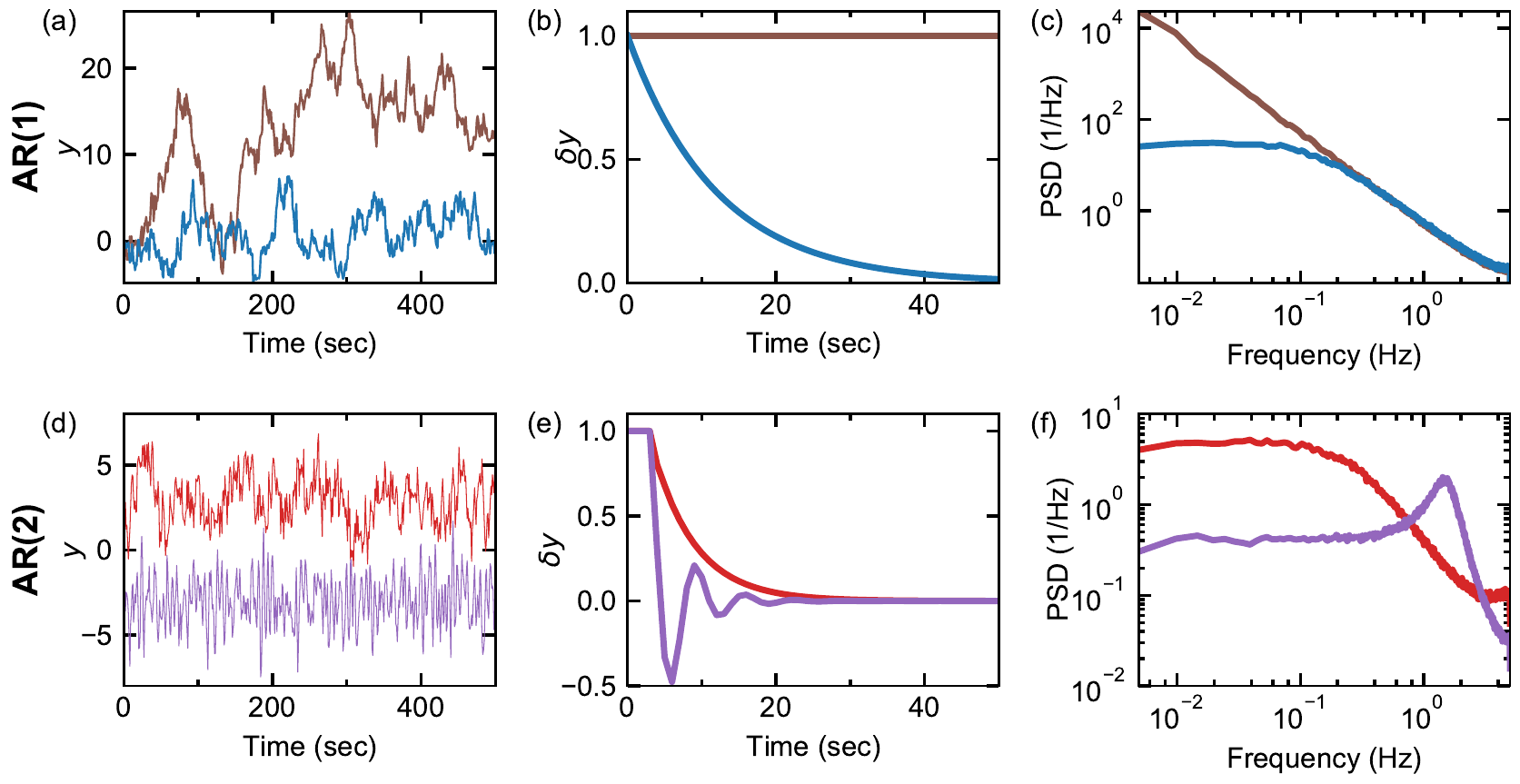}
    \caption{Examples of AR(1) and AR(2) processes. First row (a-c) shows two examples of the AR(1) process with parameters $a_1=1$, representing Brownian motion (brown line) and $a_1=0.92$ depicting a damped random walk (blue line); second row (d-f) illustrates examples of the AR(2) process with parameters $a_1=0.6, \, a_2=0.2$ (red line) having an exponential ACF, and $a_1=0.9, \, a_2=-0.6$ (purple line) featuring an oscillation around the mean value. From left to right, the panels display time traces, i.e., realizations of random process, propagations of a one-time shock, and the power spectral densities of the respective processes.}
    \label{fig:ar_examples}
\end{figure*}

\subsection{Additional power spectral density data}
\label{sub:add_PSD}
In Fig.~\ref{fig:psd_compar}\,(a-c) further frequency noise PSD data are shown, confirming the discussed noise independence of power and temperature. Data for panel (a) and (b) are taken at $\omega_{\mathrm{r}} = \SI{6.11}{\giga \hertz}$, and therefore the sample is fully magnetized. Over the measured range of the two external parameters, no difference in the low-frequency part of the PSDs can be seen. Again, the increase in white noise in panel~(a) is attributed to amplifier noise due to lower input power \cite{boudot_phase_2012}. Figure~\ref{fig:psd_compar}\,(c) displays a comparison of two different resonance frequencies, above and below the saturation magnetization. The data are normalized via the resonance frequencies for better comparison. The low frequency parts are identical. Yet, a small difference is visible around the kink at \SI{1}{\hertz}, suggesting a small influence of the sample magnetization.

\subsection{Time series analysis}
\subsubsection{Examples of AR processes}
\label{sub:examples_AR}

For a better understanding of AR processes, we show examples of first and second order AR processes with different parameters $a_i$. The mathematical background can be found in a textbook by Box \textit{et al.}\cite{box_time_2015}, for instance. Recall from Eq.~\ref{eq:AR}, that in the AR process, white noise fluctuations are added to a weighted sum of the last $p$ data points. Examples of time traces generated according to this equations are displayed, as well as the propagation of a one time-shock and their PSDs are shown in Fig.~\ref{fig:ar_examples}. As mentioned, in the main text, an AR(1) process with $a_1=1$, represents the famous random walk or Brownian motion. In Fig.~\ref{fig:ar_examples}\,(a), we can see how the summing of all white noise terms leads to a few big fluctuations over time. The summation can be also seen as the integration of white noise, giving an $1/f$ term in frequency space, which is subsequently squared for the PSD and therefore yielding the $1/f^2$ decay of so-called brown noise. Reducing $a_1$ filters out big fluctuations, since a one-time shock is exponentially damped over time (compare Fig.~\ref{fig:ar_examples}\,(a,b)). This filtering is also visible in the PSD (Fig.~\ref{fig:ar_examples}\,(c)), having a Lorentzian form, which flattens out for low frequency compared to the pure random walk. Now, considering an AR(2) process and choosing the parameters accordingly, we see either the exponential damping of the white noise fluctuations for $a_1=0.6, \, a_2=0.2$ or an oscillating behavior for $a_1=0.9, \, a_2=-0.6$ (Fig.~\ref{fig:ar_examples}\,(e)). Moreover, the differences can already be recognized in their time traces (Fig.~\ref{fig:ar_examples}\,(d)), where the oscillating AR(2) process frequently crosses the mean value. In the PSD, Fig.~\ref{fig:ar_examples}\,(f), the damped process has a form close to a Lorentzian, whereas the oscillating process shows a peak at its oscillating frequency. Increasing the order of the AR process shows a similar qualitative behavior with oscillations and/or damped exponentials, but described by additional summands.

\subsubsection{Yule-Walker equations and partial autocorrelation}
\label{sub:YWE}
In the main text, we stated the usefulness of the Yule-Walker equation (YWE) to estimate the specific values for the $a_i$ and the partial auto correlation function (PACF) to estimate the order of the AR process. Again, for more mathematical derivations, we refer to Box \textit{et al.}\cite{box_time_2015} and present only the employed procedure for our calculations. The YWE are a set of equations that relate the values of the ACF $r_i$ at lag $i$ to the coefficients $a_i$ of the AR process and are defined as follows: 

\begin{equation}
\left(\begin{array}{cccc}
1 & r_{1} & r_{2} & \dots\\
r_{1} & 1 & r_{1} & ...\\
r_{2} & r_{1} & 1 & ...\\
\vdots & \vdots & \vdots & \ddots\\
r_{p-1} & r_{p-2} & r_{p-3} & \dots
\end{array}\right)\left(\begin{array}{c}
a_{1}\\
a_{2}\\
a_{3}\\
\vdots\\
a_{p}
\end{array}\right)=\left(\begin{array}{c}
r_{1}\\
r_{2}\\
r_{3}\\
\vdots\\
r_{p}
\end{array}\right).
\label{eq:yule_walker}
\end{equation}
We see that for a specific order $p$, we obtain a set of $p$ equations, in which we can replace the $r_i$ by their measured values and solve for the $a_i$.
Moreover, the PACF can also be calculated with the YWE. Since the PACF only describes the direct correlation between data points and since there is no dependence for lag values $n>p$ in the AR process, the coefficient $a_p$, equaling the order of the AR process also represents the PACF value at lag $p$. This means one has to start with order $p=1$, take the measured $r_1$, and calculate $a_1$ (which equals $r_1$) for the first value in the PACF. Then $p$ needs to iteratively be increased and the procedure repeated. If an AR process is applicable the PACF will drop to zero after the first $p$ values. The white noise part can be considered as zero order of the AR process and can also be incorporated into the Yule Walker equations as
\begin{equation}
    r_0 = \sum_{i=1}^p a_i r_i + \sigma^2 = 1.
\end{equation}
Hence, after the $a_i$ are determined, the variance of the Gaussian white noise process $\sigma^2$ can be estimated. For the numerical time-series analysis in this work, we employed the python statsmodel\cite{seabold_stats_2010} package.

\subsection{PACF dependence on sampling rate}
\label{sub:PACF_sampl}
\begin{figure}
    \centering
    \includegraphics{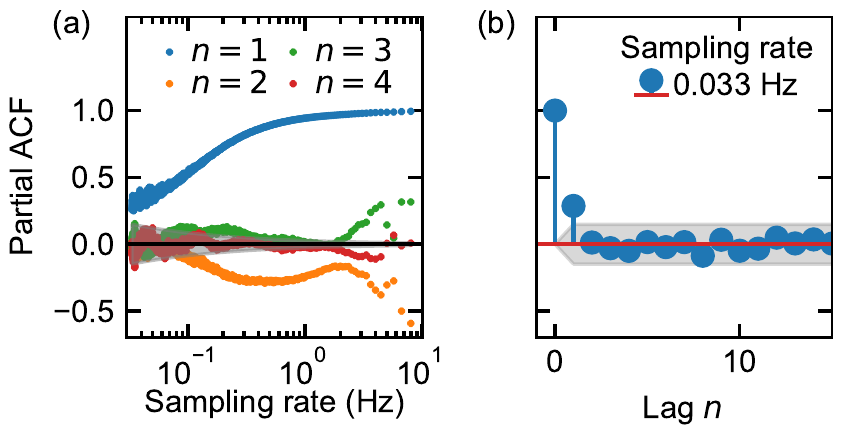}
    \caption{Dependence of the partial autocorrelation (PACF) on the post-processing sample rate. (a) PACF for lag $n=$ 1 to 4. (b) PACF at lowest sampling rate for different lags. Only first lag shows a value significantly different from zero, as depicted by the grey region, the \SI{95}{\percent} confidence interval.}
    \label{fig:PACF_bw}
\end{figure}

We showed the time series analysis for a post averaged sampling rate of \SI{8}{\hertz} in the main text. The sampling rate was chosen such that the steep decay in the PSD is still captured but the influence of the $1/f$ HEMT noise and the periodic signals, mainly \SI{50}{\hertz} current oscillations, are averaged out. Now, we consider different sampling rates below \SI{8}{\hertz}. Figure~\ref{fig:PACF_bw}\,(a) shows the PACF for the first four lags depending on the sampling rate. We see that the third order becomes negligent below \SI{2}{\hertz}, where also the steep decay is averaged away. The PACF value at lag $n=2$ remains for even lower sampling rates, likely because of the slight curvature in the PSD leading to the knee at \SI{1}{\hertz}. Reducing the sampling rate even lower, the PSD becomes a simple Lorentzian and hence only the PACF at lag $n=1$ is of importance. Values at higher lags are within the grey shaded region denoting the \SI{95}{\percent} confidence interval and are hence not significant anymore. Figure~\ref{fig:PACF_bw}\,(b) emphasizes this point by showing the PACF for several lag values at the lowest evaluated sampling rate.
\FloatBarrier

\subsection{Room temperature characterization}
\label{sub:RT_char}
\begin{figure}
    \centering
    \includegraphics{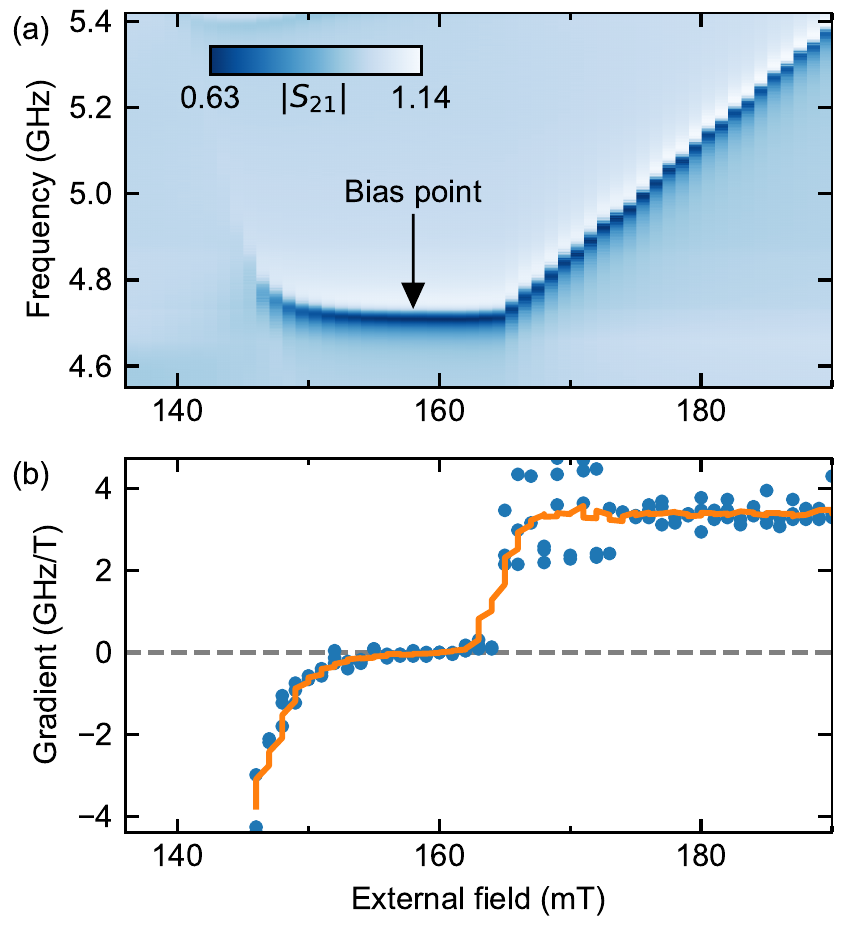}
    \caption{Dispersion relation and gradient of LiFe. (a) A flat region between \SI{150}{\milli \tesla} and \SI{165}{\milli \tesla} is visible in the dispersion relation, attributed to a mode softening \protect\cite{goryachev_cavity_2018} and making the ferromagnetic resonance less susceptible to field fluctuations. The arrow indicates the bias point at which the fluctuation measurements were performed. Values higher than one in the $S$ matrix element are due to the background correction in combination with an impedance mismatch in the system. (b) Gradient of the dispersion spectrum, numerically calculated. Points are the extracted FMR frequencies with a median filter (solid line) as a guide to the eye.}
    \label{fig:rt_disper}
\end{figure}

Figure~\ref{fig:rt_disper}~(a) shows the dispersion relation of the Kittel mode and its gradient (b) for LiFe at room temperature. Goryachev \textit{et al.} observed a minimum in the dispersion due to a mode softening\cite{goryachev_cavity_2018}. There the resonance frequency is first-order insensitive to fluctuations in the external field. For our sample, the FMR dispersion exhibits a flat region over roughly \SI{15}{\milli \tesla}. Due to the high linewidth, $\kappa \approx \SI{17}{\mega \hertz}$ (HWHM) and therefore the small slope of the phase response, frequency noise of LiFe could not be observed at low temperature. It was masked by the HEMT phase noise. We note that the linewidth at this minimum is higher than in the linear region ($\kappa\approx\SI{13}{\mega \hertz}$) and also that the sample is not fully magnetized.  
\FloatBarrier

\end{document}